\documentclass[aps,prl,twocolumn]{revtex4} 

\usepackage{psfig}

\input{macros.phys}

\begin{document}

\title{Detection of confinement and jumps
in single molecule membrane trajectories}

\author{N. Meilhac,$^1$ L. Le Guyader,$^2$ L. Salom\'e$^2$ and N. Destainville$^1$}
\affiliation{$^1$Laboratoire de Physique Th\'eorique, UMR CNRS-UPS 5152,
Universit\'e Paul Sabatier, 31062 Toulouse Cedex 9, France  \\
$^2$Institut de Pharmacologie et de Biologie Structurale, UMR CNRS-UPS 5089,
205, route de Narbonne, 31077 Toulouse Cedex, France.}

\date{\today}

\begin{abstract}
We propose a novel variant of the algorithm by Simson {\em et
al.} [R. Simson, {\em et al.},  {\em Biophys. J.} 
{\bf 69}, 989 (1995)]. Their algorithm was developed to detect
transient confinement zones in experimental single particle tracking
trajectories of diffusing membrane proteins or lipids. We show that
our algorithm is able to detect confinement in a wider class of
confining potential shapes than Simson {\em et al.}'s one. Furthermore
it enables to detect not only temporary confinement but also {\em
jumps} between confinement zones. Jumps are predicted by membrane
skeleton fence and picket models. In the case of experimental
trajectories of $\mu$-opioid receptors, which belong to the family of
G-protein-coupled receptors involved in a signal transduction pathway,
this algorithm confirms that confinement cannot be explained solely by
rigid fences.
\end{abstract}

\maketitle

One of the central issues of contemporary cellular biology is to
establish the relationship between dynamical organization and
biological functions of membrane constituents. The development of
single particle tracking (SPT) techniques gives enthusiastic new
insights into dynamical organization of membranes, that were so far
inaccessible by ensemble-average methods. Indeed, the diffusive motion
of molecules of interest (proteins or lipids) at the surface of living
cells can be followed with a nanometric resolution, 
after labeling them by means of
fluorophores, gold colloids, latex beads, or quantum
dots~\cite{Saxton97}. Yet, new specific and performant tools must be
developed to extract valuable information from these trajectories.

\begin{figure}[ht]
\begin{center}
\ \psfig{figure=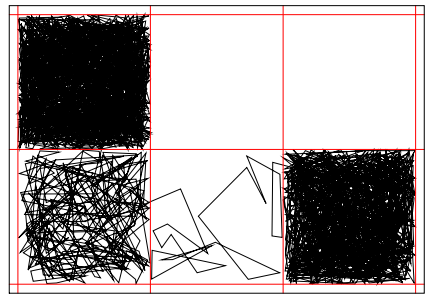,width=5.6cm}
\end{center}
\vspace{-5mm}
\caption{(Color online) Example of trajectory in a square grid of
rigid barriers of periodicity $L$.
At short times, the particle diffuses in a closed box, where
it stays on average a time $\tau_{res}$. Periodically,
it jumps from one box to an adjacent one, thus resulting in
a slow long-term diffusion with constant $D_{M}$.\label{lattice}}
\end{figure}

After more than 15 years of efforts using SPT techniques, the question
of the cell membrane organization and compartmentalization is still
the matter of intense and controversial
debate~\cite{Jacobson95,Daumas03,Choquet03,Chen04,Suzuki05,Jacobson05}. What
is the origin of the confinement quite generally observed in tracking
experiments? How is it related to the transmission of signals through
the membrane? Confinement is indeed commonly observed in SPT
trajectories: the diffusive motion is not purely Brownian, but rather
affected by either rigid obstacles, confining domains such as rafts
(or other signaling platforms), or molecular interactions. Confinement
can be transient~\cite{Chen04}, the molecule being now and then
trapped in ``transient confinement zones'' (TCZ). Experimental
situations also exist where the molecule is always confined, while
showing a long-term slow diffusion. There exist different models that
account for such a behavior. In the ``membrane skeleton fence and
picket models''~\cite{Kusumi93,Fujiwara02}, the confinement is due to
the cytoskeleton of the cell close to the membrane, or by proteins
anchored to it, which form rigid corrals. Successive hops between
adjacent domains result in a slow long-term diffusion of the molecules
(Fig.~\ref{lattice}). The recent alternative ``interacting Brownian
particle model''~\cite{Daumas03} proposes that~-- in the case studied
in this reference~-- barriers do not satisfyingly explain the observed
confinement, which more likely originates from long-range attractive
interactions between membrane proteins. The latter form auto-assembled
aggregates in which proteins are trapped. The long-term diffusive
behavior is the manifestation of the diffusion of the center of mass
of the assembly. We demonstrate that the algorithm studied in this
paper is able to discriminate between these models. Indeed, in a fence
and picket model, proteins regularly jump from a confining zone to an
adjacent one. By contrast, the ``interacting Brownian particle model''
does not require jumps to account for long-term slow diffusion. We
show that, despite statistical fluctuations, our algorithm detects
jumps with good confidence, when they exist. We calculate how many
jumps are theoretically expected in a fence and picket model, and we
compare this number to the effectively detected ones in experimental
trajectories of Ref.~\cite{Daumas03} ($\mu$-opioid receptors). We find
that there is an unequivocal discrepancy. This proves that rigid
fences cannot be considered as the unique source of confinement.

Beyond this particular example, our algorithm intends to be applicable
to a wide range of experimental situations. It responds to an
increasing demand consecutive to the rapid development of SPT
experiments. It is a challenge to develop a simple and reliable tool
to discriminate between different sources of confinement or more
simply between confined and non-confined trajectories. The present
Letter intends to propose such a robust tool.

Simson {\em et al.}'s algorithm~\cite{Simson95} has been designed to
detect transient confinement. It is based on the following principle.
Consider a Brownian trajectory $\vect{r}(t)$ on a time interval
$[t_0,t_0+\delta t]$. The maximum of $|| \vect{r}(t) - \vect{r}(t_0)
||^2$ on $[t_0,t_0+\delta t]$, denoted by $\maxindex{r}^2$, scales
like $D \delta t$ where $D$ is the diffusion constant. Then the
authors define a confinement index $\lambda$ (denoted by $L$ in their
paper) that is an affine function of $D \delta t/\maxindex{r}^2$.  $D$
is determined by measuring the slope at the origin of the mean square
deviation MSD$(t)$. If the diffusion is confined in a domain of
typical size $L$, then $\maxindex{r}^2$ is limited by $L^2$,
and $\lambda$ is larger than in the case of free Brownian diffusion. The
authors determine a threshold $\lambda_c$.  Roughly speaking, if $\lambda>\lambda_c$,
the diffusion is confined, otherwise it is free (see
Ref.~\cite{Simson95} for more details). Along a trajectory, $\lambda(t)$ is
calculated on sliding intervals $[t-\delta t/2,t+\delta t/2]$. The
plot of $\lambda(t)$ indicates TCZs as intervals where $\lambda > \lambda_c$.

\begin{figure}[ht]
\begin{center}
\begin{tabular}{l}
\ \vspace{-0.17cm}\psfig{figure=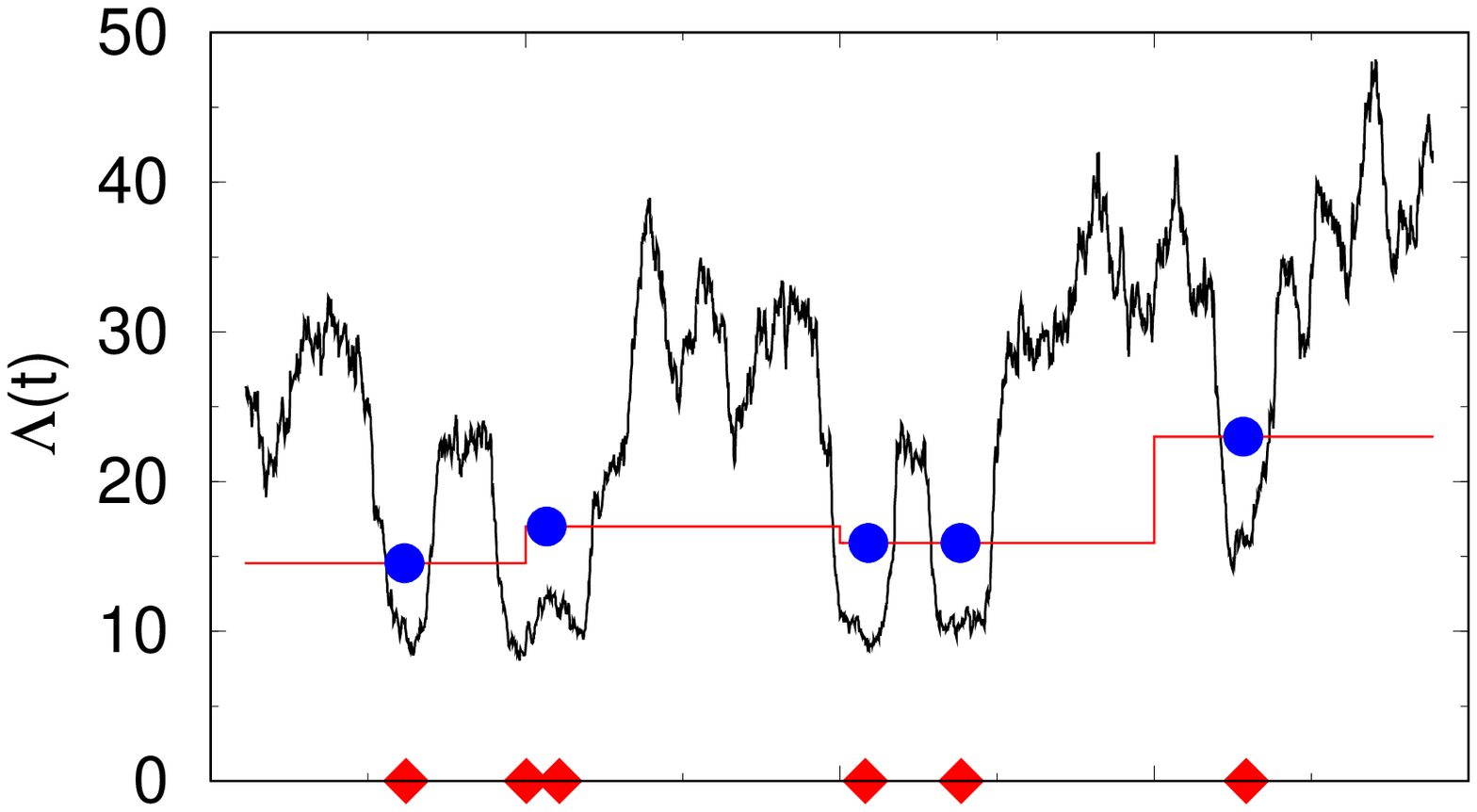,width=6.73cm} \ \\ 
\ \vspace{-0.7cm}\psfig{figure=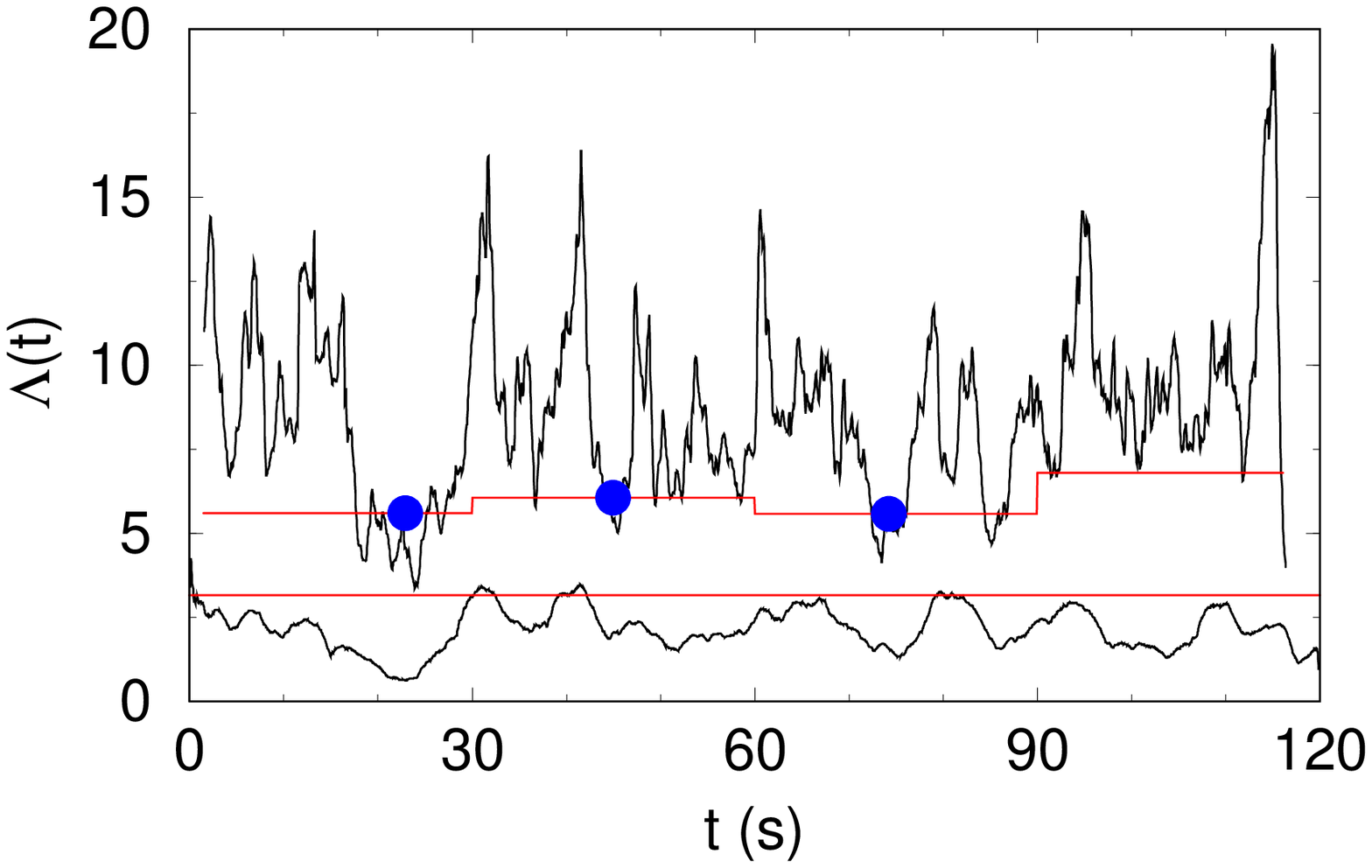,width=7cm} \
\end{tabular}
\end{center}
\caption{(Color online) Top: Profile $\Lambda(t)$ for a numerical
trajectory with $\tau_{res}=20$~s. We have also plotted the
threshold $\alpha \overline{\Lambda}(t)$ with $\alpha=0.7$. Real jumps
are represented by diamonds and detected ones by circles. One can see
5 detected jumps among which one double jump (detected only
once). Note that the first and last $\delta t/2$ segments are not
taken into account because $\Lambda(t)$ cannot be calculated
there. Bottom: Upper plot: Profile for an experimental trajectory,
with 3 detected jumps. Here $\tau_{res}=9.6$~s and the fence model
would predict 12 jumps. Some intervals where $\Lambda(t) \leq \alpha
\overline{\Lambda}(t)$ are not considered as jumps because they are
not long enough. Lower plot: Profile $\lambda(t)$ calculated with
Simson et {\em al.}'s algorithm~\cite{Simson95} and the threshold
$\lambda_c=3.16$ (see~\cite{Simson95} for details). The receptor is
apparently hardly ever confined.
\label{profils} }
\end{figure}

We show that even though Simson {\em et al}'s method is applicable to
a large variety of experimental cases, there exist situations of great
interest where it is not operational. First, artifactual detections of
TCZs can happen when $D$ varies along the
trajectory~\cite{Daumas03}. This problem is fixed by computing $D$
locally by the same method, on intervals of duration a few
seconds. More importantly, this method fails in detecting confinement
in non-flat potentials. For example, in a quadratic well of typical
width $L$ at temperature $T$, the molecule is likely to explore
regions of energy of several $k_B T$ where $r \gg L$, and the measured
$\maxindex{r}^2$ fluctuates a lot around its typical value, depending
on whether such rare points are in the trajectory or not. We
experienced that it happens that Simson {\em et al.}'s algorithm does
{\em not} detect flagrant confinement in quadratic potentials or in
experimental trajectories (see Fig.~\ref{profils} for an example). The
algorithm proposed in Ref.~\cite{Kusumi05}, also based on
$\maxindex{r}^2$, presents the same limitations. This is intrinsic to
the methods, namely the choice of $\maxindex{r}^2$ to characterize
trajectory wanders, and not to a particular choice of parameters.

For this reason, we modify the above algorithm as follows: instead of
calculating $\maxindex{r}^2$, we compute the variance $\Delta r^2(t)$
of $\vect{r}$ on intervals $[t-\delta t/2,t+\delta t/2]$. Rare points wandering far
away from the potential minimum thus have a low weight in $\Delta r^2(t)$,
which gives a more accurate measurement of the typical width of the
confining potential. Of course, in the case of flat potentials
delimited by rigid fences, both algorithms present a similar
efficiency. A novel confinement index is now defined 
\begin{equation}
\label{level}
\Lambda = \frac{D\delta t}{\Delta r^2}.
\end{equation}
Up to a numerical prefactor, $\Lambda$ is the ratio of the variance of
a free random walk to the one of the trajectory under study. If it is
unconfined, $\Lambda$ will be of order unity, whereas it will be large
in the converse case. We calculate the typical values of $\Lambda$ in
the respective cases of free Brownian two-dimensional trajectories and
confined ones.  We model our Brownian molecule by an over-damped
Langevin particle: $d\vect{r}/dt = \vect{\eta}(t)$, where
$\vect{\eta}$ is a Gaussian white noise: $\langle \vect{\eta} \rangle=
\vect{0}$ and $\langle \eta_i(t) \eta_j(t') \rangle = 2D \delta_{ij}
\delta(t-t')$. Then the mean position $\overline{\vect{r}}$ and the
mean square position $\overline{\vect{r}^2}$ are calculated on a {\em
single} trajectory before being averaged over noise, leading to
$\Delta r^2 = \langle \overline{\vect{r}^2} - \overline{\vect{r}}^2
\rangle = (2/3) D\delta t$.  Statistical fluctuations of this
quantity can also be calculated using Wick's theorem~\cite{Wick} to
compute 4-time correlators of $\vect{\eta}$: $\Delta \langle
\overline{\vect{r}^2} - \overline{\vect{r}}^2 \rangle =
\frac{2\sqrt{2}}{3 \sqrt{5}} D\delta t$. Thus $\Lambda_{Brown}
\lesssim 1/(2/3-2\sqrt{2}/3 \sqrt{5}) \simeq 4$ for a pure Brownian
trajectory, independently of $\delta t$ and $D$, as checked on
numerical trajectories. Now we consider confined diffusion in a square
box of side $L$.  If one averages over $N \gg 1$ {\em independent}
positions, one gets $\Delta r^2=L^2/6$,
\begin{equation}
\label{lambda_conf}
\Lambda_{conf} = 6D\delta t / L^2.
\end{equation}
In this confined case, the statistical fluctuations of $\Delta r^2$
vanish at large $N$. Hence TCZs will be distinguishable from pure
Brownian trajectories if $\Lambda_{conf} \gg \Lambda_{Brown}$ {\em
i.e.} if $\Lambda_{conf} \gg 4$ or $D\delta t/L^2 \gg 2/3$. As
expected, $\delta t$ must be sufficiently large to enable
detection. Note that above we have asked for the number of {\em
independent} images to be sufficiently large. In practice, positions
in successive images {\em are} correlated, because the equilibriation
time to explore a box of side $L$ is $\tau = L^2/\pi^2
D$~\cite{Daumas03}.  One must make sure that $N \gg 1$ with respect to
this time.

Our new algorithm detects confinement with excellent reliability,
whatever the shape of the box or of any non-flat confining potential.
Now how does it detects jumps over fences? We suppose first that the
molecule evolves in a grid of square boxes of side $L$, separated by
rigid barriers, and that it can jump over a barrier by thermal
activation (Fig.~\ref{lattice}). It has a long-term slow diffusion
with a constant $D_{M}$: $\Delta r^2 = 4D_{M}\; t$ at large $t$. If
$\tau_{res}$ is the average time of residence in a box, {\em i.e.} the
average time between successive jumps, then $L^2 = 4 D_{M}
\tau_{res}$. When the molecule is confined in a single box, $\Lambda =
\Lambda_{conf}$. If there is a jump at time $t_0$ in a segment
$I=[t-\delta t/2,t+\delta t/2]$, the particle is virtually in a larger
box $2L \times L$: it spends a time $t_0 -t + \delta t/2$ in a box and
the remaining time in an adjacent one. If $\delta t \gg \tau$, the
probability distribution of the molecule position is close to uniform
in each box. This permits to calculate $\Delta r^2$ by partitioning $I$
in two intervals, one for each box. One gets
\begin{equation}
\label{lambdaTCZ}
\Lambda(t) = \frac{2}{5}\ \frac{\Lambda_{conf}}{1 - \frac{12}{5}
\frac{(t-t_0)^2}{\delta t^2}} < \Lambda_{conf}.
\end{equation}
There is a gap centered at $t_0$ in the profile $\Lambda(t)$ and
$\Lambda(t_0) = 2\Lambda_{conf}/5 $ (see Fig.~\ref{profils}). In the
case of multiple jumps in $I$, the corresponding gaps
merge. To resolve single jumps at best, we choose $\delta t \leq
\tau_{res} / 3$. In addition, $\delta t$ must be as large as possible
to get higher profiles $\Lambda(t)$ where confinement is best
detected. Therefore we set
\begin{equation}
\label{Smax}
\delta t = \frac{1}{3} \; \tau_{res} = \frac{L^2}{12\; D_{M}}.   
\end{equation}

In order to detect these gaps, one must also make sure that their
minima are higher than $\Lambda_{Brown}$, the ``background noise'' of
pure Brownian trajectories. Indeed, if not, the depth of the gap will
be reduced, and this will corrupt detection. 
This condition reads $2\Lambda_{conf}/5 > 4$, or $\Lambda_{conf}>10$.
Together with Eq.~(\ref{Smax}), it can be written
in terms of the measurable quantities $D$ and $D_{M}$:
\begin{equation}
\label{cond1}
D > 20\; D_M.
\end{equation}
In other words, long-term and short-term time scales must be well
separated. This is observed in a large majority of the experimental
trajectories below.

Now we evaluate the capabilities of our algorithm on numerical
trajectories. They simulate Brownian molecules evolving in a square
grid of rigid barriers of periodicity $L$. When a step crosses a
barrier, it is allowed with a probability $p$ suitably defined so that
the long-term diffusion constant equals $D_{M}$. Numerical parameters
match those of experimental trajectories (see below). One image is
sampled every 40~ms. In addition, $D$ is allowed to vary slowly with
time in a given trajectory, to mimic possible composition or physical
changes of the underlying membrane.  More precisely, every second, $D$
is multiplied by a factor randomly chosen in the interval
$[0.9,1.1]$. To calculate $\Lambda$, we {\em measure} the diffusion
constant $D_m$ by calculating the MSD on intervals of 5~s and fitting
the slope at the origin (first 3 points).  We check that $D_m \simeq
D$. Because of statistical fluctuations on finite samples, all jumps
cannot be detected and there are false detections. We denote by
$\sigma$ the fraction of jumps successfully detected by the algorithm
among real jumps and by $\overline{\sigma}$ the ratio of false
detections to real jumps.  The higher $\sigma$ and the lower
$\overline{\sigma}$, the best the algorithm.  To localize jumps, we
need to estimate the value $\Lambda_{conf}$ since gaps are intervals
where $\Lambda(t)$ is significantly smaller than the confinement value
$\Lambda_{conf}$ for a sufficient duration. In practice, we proceed as
follows. We compute $D_m$ as detailed above. We measure $\Delta r^2$
and we calculate $\Lambda(t)$ (see Fig.~\ref{profils}). To avoid
biases due to the slow variations of $D$, we calculate the average
$\overline{\Lambda}(t)$ of $\Lambda(t)$ over successive segments of
30~s; $\overline{\Lambda}(t)$ is our estimation of the confined
profile: $\overline{\Lambda}(t) \simeq \Lambda_{conf}$, because we
anticipate that jumps are rare.  Next we detect intervals where the
signal is well below $\overline{\Lambda}$. More precisely, we require
that $\Lambda(t) \leq \alpha \overline{\Lambda}(t)$ for a duration
larger than $t_c$, where the parameters $t_c$ and $\alpha \in [0,1]$
must be optimized. We have scanned large ranges of values of both
$\alpha$ and $t_c$, and calculated $\sigma$ and $\overline{\sigma}$ in
each case ($10^3$ trajectories of $T=$120~s), with the typical
parameters of the experimental trajectories below: $D\simeq
0.1$~$\mu$m$^2$s$^{-1}$, $L\simeq 0.3$~$\mu$m, $\tau_{res} =
10-20$~s. We observed that $t_c=\delta t/2$ and $\alpha=0.7$ gives
the best compromise with $\sigma \geq 63\%$ and
$\overline{\sigma} \leq 0.7 \% $. The algorithm detects two thirds of
the jumps and makes very few false detections. The value of $\sigma$
comes from the fact that close jumps cannot be resolved and are
counted only once.
%Note that if needed, true and false detections can be
%directly discriminated ``by eyes'' on trajectory plots with good
%confidence, since the algorithm determines at which time they are
%supposed to happen.

Another serious complication can arise in experimental
trajectories. The confinement domains are not necessarily squares.  If
they are elongated, like rectangles or more complex shapes,
Eq.~(\ref{lambdaTCZ}) is no longer valid. Consider for instance
rectangles $L \times \rho L$. There are two types of jumps, 
over short or long edges. It can easily be
quantified how this affects the relative depths of the gaps associated
with each type of jump.  If the rectangle is extremely elongated, then
only jumps over short edges can be reasonably detected. This
complication can be overcome by measuring $\Delta x^2$ and $\Delta
y^2$ and multiplying the latter by a counterweight: $\Delta_{cw} r^2 =
\Delta x^2 + \Delta y^2/\rho^2$. Then both kinds of gaps again have
the same depth and the previous analysis becomes valid. If the main
axis of the box are not parallel to the axes $Ox$ and $Oy$, then
before applying counterweights, one must recover the average
directions of these main axes by diagonalizing the correlation matrix
$C=\langle \vect{r}(t)\vect{r}^T(t) \rangle - \langle
\vect{r}(t)\rangle\langle\vect{r}^T(t) \rangle$ averaged over
sufficiently long time intervals (typically 30~s).  We checked that
this procedure is operational, even though it increases significantly
the number of false detections because of additional numerical
operations. However, this question goes beyond the scope of this
Letter because in the experimental trajectories below, by
diagonalizing $C$, we find $\rho \simeq 1.4$ on average (while
$\rho \simeq 2.1$ for a pure unconfined random walk~\cite{rudnick86}), in
which case both types of gaps have typically the same depth
and $\sigma$ and $\overline{\sigma}$ are not significantly affected if
one uses the original profile~(\ref{level}): $\sigma \geq 55\%$ and
$\overline{\sigma} \leq 2.5\%$.

Now we apply our algorithm to the
102 experimental trajectories from Ref.~\cite{Daumas03}.  These are
trajectories of $\mu$-opioid receptors at the surface of normal rat
kidney fibroblast cells, tracked by SPT, after being labeled by 40~nm
gold colloids, at 40~ms time resolution during $T=$~120~s. The
parameters $D$, $D_{M}$ and $L$ are
measured~\cite{Daumas03} by fitting the MSD by $MSD(t) = (L^2/3) (1
- \exp(- 12Dt/L^2)) + 4 D_{M} \; t$.  Typically, $D\sim
0.1$~$\mu$m$^2$s$^{-1}$, $L\sim 0.3$~$\mu$m, $\tau_{res} \sim
10$~s. From $D_{M}$ and $L$ we deduce
$\tau_{res}$. In~\cite{Daumas03}, it was noticed that about 15\% of
the trajectories show significant slow variations of $D$ and $L^2$
in a same trajectory, up to one order of magnitude for $D$. This could
be a serious concern, because variations of these parameters cause
variations of the reference value $\Lambda_{conf}$ that could be
misinterpreted as jumps.  Fortunately, it was also noticed that on
individual trajectories, $D \propto L^2$ when $D$ and $L$ vary
along a trajectory.  Consequently $\Lambda_{conf}$ varies only
moderately (see Fig.~\ref{profils}, bottom). This is well mimicked by
the slow variations of $D$ at fixed $L$ that we have imposed in
numerical trajectories.  If $\delta t=\tau_{res}/3$ exceeds 15s, we set
$\delta t=15$~s, not to loose too many points at the beginning and the end of
the trajectory (see Fig.~\ref{profils}). We eliminate the 18
trajectories that do not satisfy condition~(\ref{cond1}), as well as
those that were qualified of ``slow or directed diffusion'' in
Ref.~\cite{Daumas03}, because their MSD were more correctly fitted by
the corresponding theoretical equations. We are left with 67
trajectories.

\begin{figure}[ht]
\vspace{-3mm}
\begin{center}
\ \psfig{figure=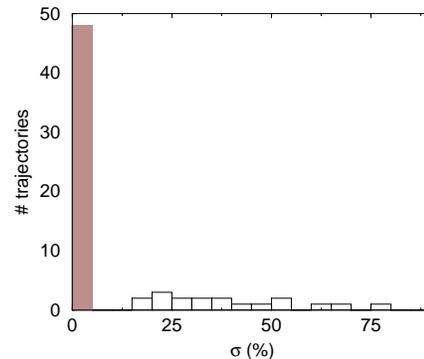,width=5.6cm} \
\end{center}
\vspace{-10mm}
\caption{(Color online) Histogram: ratios of detected jumps to the ones
expected in a fence model, on experimental 67 trajectories of
$\mu$-opioid receptors~\cite{Daumas03}. An event at $\sigma=400$\% is
not shown.
\label{Histo} }
\end{figure}

First of all, we check that $\overline{\Lambda}\gg 4$ on all
profiles. This confirms that all trajectories are confined. From the
value of $\tau_{res}$, we estimate the expected number of jumps, {\em
if the trajectories were correctly described by a fence or corral
model}, namely $(T-\delta t)/\tau_{res}$. Then we count the detected
jumps. An example is provided in Fig.~\ref{profils} (bottom). We find that the
average ratio of detected jumps to the ones expected with fence or
corral models is only $\sigma_{exp}=16.4$\%, where we expected more
than 55\%. The histogram in Fig.~\ref{Histo} gives greater
details. Therefore, as already concluded using independent arguments
in Ref.~\cite{Daumas03}, a fence or corral model {\em is not able to account
alone for experimental observations}.

We clearly see that two populations of receptors emerge in the
histogram. The first population (empty bars) contains
19 trajectories, of average detection ratio $\sigma'_{exp}=58$\%, the
detected jumps of which can perfectly be accounted for by a fence or
picket model.
%Let us focus on the 
%trajectories where no jumps at all are detected. They can be
%divided into two subgroups. In 28 trajectories, $\tau_{res}<120$~s and
%we expect 2 jumps on average in a fence or picket model. In the 21
%remaining ones, $\tau_{res}>120$~s and we are not sure to detect any
%jump in this case. But the fence or picket model cannot explain the
%positive value of $D_{M}$ in this case since jumps are compulsory to
%get long-term diffusion. 
The second population (left shaded bar) concerns 48 trajectories where
{\em we do not detect any jump}. Their long-term diffusion {\em
cannot} be explained by any fence or picket model.

\vspace{-3mm}
We are led to the following original conclusion: hop diffusion
probably exists in trajectories of $\mu$-opioid receptors, and it
can satisfyingly explain the long-term diffusive behavior of nearly
30~\% of the analyzed trajectories. But another mechanism must be
invoked to explain long-term diffusion in a majority of cases. This
reinforces the need for an alternative model to account for long-term
diffusion, as proposed in Ref.~\cite{Daumas03}. We might even ask if
two distinct mechanisms are not independently at work in cells to
achieve confinement, and if there would not exist two populations of
$\mu$ receptors: the first ones confined by fences or pickets,
and the other ones by long-range inter-protein interactions.
Additional experiments will be necessary to test this hypothesis.

We express our gratitude to Ken Jacobson for helpful discussions. We
also acknowledge the French Ministry of Research and the CNRS for
financial support.

\end{document}